\documentclass[preprint,aps,showpacs]{revtex4-1}
\usepackage{graphicx}
\usepackage{hyperref}
\usepackage{epstopdf}
\usepackage{ulem}

\def\be{\begin{eqnarray}}
\def\ee{\end{eqnarray}}

\def\bc{\begin{center}}
\def\ec{\end{center}}

\newcommand{\bfq}{{\bf q}_{\perp}}

\newcommand{\bfk}{{\bf k}_{\perp}}
\newcommand{\bfr}{{\bf r}_{\perp}}

\newcommand{\bfb}{{\bf b}_{\perp}}

\usepackage{color}

\begin{document}

\title{Charge and longitudinal momentum distributions in transverse coordinate space} 

\author{Chandan Mondal$^1$, Narinder Kumar$^{1,2}$, Harleen Dahiya$^2$, Dipankar Chakrabarti$^1$}
\affiliation{$^1$Department of Physics, Indian Institute of Technology Kanpur, Kanpur-208016, India.\\
$^2$Department of Physics, Dr. B.R. Ambedkar National
Institute of Technology, Jalandhar-144011, India.}

\date{\today}

\begin{abstract}
We investigate the charge distributions for the $u$ and $d$ quarks in transverse coordinate space in a
light-front quark-diquark model for the nucleons using the overlaps of the wave functions constructed from the soft-wall AdS/QCD prediction. We have also obtained the charge distributions for proton
and neutron in transverse coordinate space and compared it with the distributions obtained in
impact-parameter space.
Further, we study the longitudinal momentum distributions using the wave functions in the transverse
coordinate space. We have also shown the explicit fermionic and bosonic contributions for different struck
$u$ and $d$ quarks.
\end{abstract}

\maketitle
\section{Introduction}
Hadronic structure and their properties, being nonperturbative in nature, are difficult to
understand from the first principle of Quantum Chromodynamics (QCD). In the last decade, there have been
numerous attempts to gain insight into the hadronic structure by studying the QCD inspired models
\cite{petrov,pentt,chodos,song}.
The quark-diquark model \cite{schmidt} is one of the most successful QCD inspired models to
investigate various aspects of hardonic properties where the nucleon is
considered to be a bound state of a single quark and a scalar or vector diquark state. Recently, a
light-front scalar quark-diquark model for the nucleons inspired by soft-wall AdS/QCD has been
proposed \cite{Gut} and extensively used to investigate and reproduce many interesting properties of the nucleons \cite{CM4,chakrabarti,CM6,Maji,CM_few,Chakrabarti:2016lod}. The light-front wave functions (LFWFs) in this model
are obtained by matching the electromagnetic form factors of the hadrons in the soft-wall model of
AdS/QCD which has been successful in explaining various hadronic properties, for example, hadron mass spectrum, Parton Distribution Functions (PDFs), Generalized Parton Distributions (GPDs), meson and nucleon form
factors, transverse densities, structure functions etc. \cite{BT_new3,AC,ads1,ModelII,ads2,AC4,BT1,BT1b,BT2,Rad,vega,CM,CM2,CM3,CM5,HSS,abidin08,kumar,reso}.

The AdS/CFT correspondence~\cite{maldacena} between the string theories of gravity in the AdS space
and conformal gauge field theories in the physical space-time provides a completely new set of tool
for studying the dynamics of QCD. One can represent the strong interactions of quarks and gluons
by a semi-classical gravity theory i.e., without quantum effects such as particle creation and
annihilation in higher dimensions. Even though, a perfect string theory dual of QCD is not
yet known, the AdS/CFT correspondence can still provide remarkable insight into various properties/features of
QCD including color confinement, qualitative explanation for meson and baryon spectra and
wave functions describing the hadron structure.
QCD is conformal in the ultraviolet (UV) region, whereas, in the infrared (IR) region, the confining gauge
theory with a mass gap is characterized by the scale $\Lambda_{QCD}$ and a well defined spectrum of
hadronic states. Light-front holography represents a remarkable connection between the AdS space and the light-front framework where the wave amplitude propagating in the AdS space is mapped into the
light-front wave functions of hadrons in space-time in terms of a light-front transverse variable $\zeta$ (giving the separation of
the quarks and gluonic constituents within the hadron) \cite{brodsky1,brodsky2}.

The electromagnetic form factors, probed through elastic scattering, contain information about the internal structure of the nucleons. A Fourier transformation  of these form factors provides information about the spatial
distributions of charge. The well known examples include the charge and magnetization
distributions inside the nucleons. One can obtain the Dirac (charge) and Pauli (magnetic) form
factors from the overlap of the LFWFs by calculating the matrix element of
electromagnetic current $J^+$ in the light-front frame \cite{BD,BHMI}. In such a frame, the momentum
transfer is the conjugate variable of the impact-parameter and via Fourier transform with respect to the momentum
transfer in transverse direction, the form factors reflect the charge distributions of quarks in
the impact-parameter space
\cite{miller07,vande,weiss,selyugin,CM4,Burkardt:2000za,burk,miller09,miller10,venkat}. The charge distributions in the transverse coordinate space obtained form the overlap of LFWFs in coordinate space have been
studied \cite{hwang,kumar} and a comparison of charge distributions in transverse impact-parameter space in different AdS/QCD models
has been reported in Refs. \cite{CM3,CM_few}.

 The gravitational form factors related to the energy momentum tensor $(T^{\mu \nu})$  also play an important role in the understanding of hadronic physics. They have been
studied in both the light-cone and the AdS/QCD framework and can also be obtained from the second moments
of GPDs \cite{kumar1,liu,chakrabarti,muller,ji,CM5,AC5,BTgrav}. The gravitational form factors
$A(Q^2)$ and $B(Q^2)$ can be obtained from helicity non-flip and helicity flip matrix elements of
the $T^{++}$ tensor current similar to the  Dirac $(F_1(Q^2))$ and Pauli $(F_2(Q^2))$ form
factors respectively.
The transverse spin sum rule and gravitational form factors have been studied \cite{chakrabarti}. The Fourier transform of the gravitational form factor $A(Q^2)$ in the
impact-parameter space has an interesting interpretation \cite{abidin08,selyugin} as it gives the longitudinal momentum density ($p^+~ density$) in the transverse impact-parameter space. A comparative study of longitudinal
momentum densities in transverse impact-parameter space in two different soft-wall AdS/QCD models
has been carried out \cite{CM5}.

The authors in Ref. \cite{hwang} had introduced the charge distribution in transverse coordinate space and explicitly calculated the distribution considering a light-front quark-diquark model where they had used the modified wavefuntions of QED Yukawa model. Similar to charge distribution in transverse coordinate space, the distribution of longitudinal momentum ($p^+$ distribution) in the transverse coordinate space can also be investigated. It would be interesting to evaluate these distributions in the AdS/QCD framework which is one of the most successful nonperturbative tools to study the hadronic properties. In the present work, we study the charge as well as the longitudinal momentum
distributions in the transverse coordinate space in a light-front quark-diquark model inspired by
soft-wall AdS/QCD and compare the consequences with the results obtained in \cite{hwang}. We take the phenomenological light-front quark-diquark model
proposed by Gutsche {\it et. al} \cite{Gut} with the parameters given in Ref. \cite{CM6}. In this
model, the LFWFs for the proton are constructed from the two particle wave
functions obtained in soft-wall AdS/QCD \cite{BT2}. The parameters in this model are fixed by fitting to the electromagnetic form factors of the
nucleons. To obtain the charge distribution in coordinate space, we have taken the Fourier transform of LFWFs in momentum space. These wave
functions have been used to obtain the charge distribution in coordinate space. Using the charge
and isospin symmetry, we have also calculated the charge distributions for proton and neutron. We
have also presented the result for longitudinal momentum distribution for $u$ and $d$ quarks. We have considered the different struck quarks to obtain longitudinal momentum distributions. Explicit results of the fermionic and bosonic contributions for longitudinal momentum distributions in coordinate space have been presented. It is important to mention here that even though both $r$ and $|b_\perp|$ are conjugate to momentum $k$ and momentum transferred $\Delta_\perp$ respectively, still the density in impact parameter space cannot be interpreted as coordinate space density. The quark-scalar diquark model has been shown to reproduce many interesting nucleon properties \cite{Gut,CM4,chakrabarti,CM6,Maji,CM_few,Chakrabarti:2016lod},  but since only scalar diquark is considered  in this  model,  the model might not reproduce correctly the  nucleon observables  involving orbital angular momentum and spin of the quarks and diquarks.

The paper is organized as follows. In Section \ref{model}, a brief introduction about the nucleons
LFWFs of a quark-diquark model in AdS/QCD is given. We discuss the charge distributions in
transverse coordinate space for the $u$ and $d$ quarks as well as the nucleons in Section
\ref{charge_coordinate}.  In Section \ref{charge-density}, we present the relation between the charge
distributions in transverse coordinate space and the impact-parameter space.  The
results of longitudinal momentum distributions in transverse coordinate space for proton and the
quarks and diquarks contributions to the distributions for different struck quarks $u$ and $d$ are presented in Section \ref{longi_coordinate}. Finally, we summarize our work in Section \ref{summary}.

\section{Light-front quark-diquark model constructed by AdS/QCD}\label{model}

Here we consider a light-front quark-diquark model for the nucleons \cite{Gut} where the LFWFs
are modeled from the soft-wall AdS/QCD solution. In this model, one can contemplate the three valence quarks of the nucleons as an effectively composite system
composed of a fermion (quark) and a composite state of diquark (boson) based on one loop quantum
fluctuations.
The Dirac and Pauli form factors for quarks in this model can be evaluated in terms of overlap of
the LFWFs \cite{BD,BHMI} as
\be\label{Dirac_FF}
F_1^q(Q^2) &=& \int_0^1dx \int
\frac{d^2\bfk}{16\pi^3}~\Big[\psi_{+q}^{+*}(x,\bfk')\psi_{+q}^+(x,\bfk)
+\psi_{-q}^{+*}(x,\bfk')\psi_{-q}^+(x,\bfk)\Big],\\
F_2^q(Q^2) &=& -\frac{2M_n}{q^1-iq^2}\int_0^1dx \int \frac{d^2\bfk}{16\pi^3}~
\Big[\psi_{+q}^{+*}(x,\bfk')\psi_{+q}^-(x,\bfk)\nonumber\\
&+&\psi_{-q}^{+*}(x,\bfk')\psi_{-q}^-(x,\bfk)\Big],
\ee
with $\bfk'=\bfk+(1-x)\bfq$ for the struck quark. Here $x$ is the light-cone momentum fraction and $\psi_{\lambda_q q}^{\lambda_N}(x,\bfk)$ are the LFWFs with nucleon helicities $\lambda_N=\pm$ and for the struck quark $\lambda_q=\pm$, where plus and minus
correspond to $+\frac{1}{2}$ and $-\frac{1}{2}$ respectively. In the frame $q=(q^+, q^-, \bfq)=(0,0,\bfq)$, we have
$Q^2=-q^2=\bfq^2$. The LFWFs defined at an initial scale $\mu_0=313$~MeV \cite{Gut} are given as
\be\label{WF}
\psi_{+q}^+(x,\bfk) &=&  \varphi_q^{(1)}(x,\bfk) \,,\nonumber\\ 
\quad
\psi_{-q}^+(x,\bfk) &=& -\frac{k^1 + ik^2}{xM_n}   \, \varphi_q^{(2)}(x,\bfk) \,, \nonumber\\
\psi_{+q}^-(x,\bfk) &=& \frac{k^1 - ik^2}{xM_n}  \, \varphi_q^{(2)}(x,\bfk)\,, \\
\psi_{-q}^-(x,\bfk) &=& \varphi_q^{(1)}(x,\bfk),\nonumber
\ee
where $\varphi_q^{(i=1,2)}(x,\bfk)$ are the modified wave functions constructed by soft-wall
AdS/QCD, after introducing the parameters $a_q^{(i)}$ and $b_q^{(i)}$ for quark $q$ \cite{Gut} and are defined as
\be
\varphi_q^{(i)}(x,\bfk)&=&N_q^{(i)}\frac{4\pi}{\kappa}\sqrt{\frac{\log(1/x)}{1-x}}x^{a_q^{(i)}}
(1-x)^{b_q^{(i)}}
\exp\bigg[-\frac{\bfk^2}{2\kappa^2}\frac{\log(1/x)}{(1-x)^2}\bigg].
\ee
For $a_q^{(i)}=b_q^{(i)}=0$, $\varphi_q^{(i)}(x,\bfk)$ reduces to the AdS/QCD solution
\cite{BT2}. In this work, we have taken the AdS/QCD scale parameter $\kappa =0.4$ GeV, obtained by
fitting the nucleon form factors in the AdS/QCD soft-wall model \cite{CM,CM2}. All the
parameters $a^{(i)}_q$ and $b^{(i)}_q$ with the constants $N^{(i)}_q$ are determined by
fitting the electromagnetic properties of the nucleons: $F_1^q(0)=n_q$ and $F_2^q(0)=\kappa_q$
where the number of valence $u$ and $d$ quarks are $n_u=2$ and $n_d=1$ in proton and
the anomalous magnetic moments for the $u$ and $d$ quarks are $\kappa_u=1.673$ and
$\kappa_d=-2.033$ \cite{CM6} . One can write the flavor decompositions of the Dirac and Pauli form factors of
nucleons in a straightforward manner as
\be\label{N_ff}
F_i^{p(n)}=e_u F_i^{u(d)}+e_d F_i^{d(u)},~~~~~~~~(i=1,2)
\ee
where $e_u$ and $e_d$ are the charges of $u$ and $d$ quarks in units of positron charge ($e$).
The nucleon form factors as well as the flavor form factors in this model have already been calculated in
Ref. \cite{CM_few} and are found to agree very well with the experimental data.

\section{Charge distributions in coordinate space}\label{charge_coordinate}

In order to evaluate the charge density in coordinate space, the LFWFs in the
transverse coordinate space $\tilde{\psi}(x,r_\perp)$ can be obtained by taking the Fourier
transform of the LFWFs in the momentum space $\psi(x, \bfk)$, as
\be
\tilde{\psi}(x, \bfr)= \int \frac{d^2 \bfk}{(2 \pi)^2} e^{i \bfk \cdot \bfr} \psi(x, \bfk).
\ee
The charge distribution in transverse coordinate space for a particular flavor in the quark-diquark model is
defined as \cite{hwang}
\be \label{defi}
P_{f}^q(r)&=& \int dx P_{f}^q(x,r) = \int dx
\left[\tilde{\psi}^{+*}_{+q}(x,\bfr)\tilde{\psi}^{+}_{+q}(x,\bfr) +
\tilde{\psi}^{+*}_{-q}(x,\bfr)\tilde{\psi}^{+}_{-q}(x,\bfr)\right],
\ee
where $r=|\bfr|$. To obtain the charge distribution in transverse coordinate space in the
light-front diquark model where the two-particle wave function is modeled from the
soft-wall AdS/QCD, we first calculate the LFWFs in transverse coordinate space via Fourier
transform of the wave functions given in Eq.(\ref{WF})
\be
\tilde{\psi}^{+}_{q+}(x,\bfr)&=& \int \frac{d^2 \bfk}{(2 \pi)^2} e^{i \bfk \cdot \bfr}
\psi^{+}_{q+}(x, \bfk)\nonumber\\
&=& \frac{N^{1}_{q}}{2 \pi} \  \frac{4 \pi}{\kappa} \sqrt{\frac{\log(1/x)}{1-x}}
x^{a_{q}^{1}} (1-x)^{b_{q}^{1}} \frac{e^{-r^2/4g}}{2 g}, \nonumber\\
\tilde{\psi}^{+}_{q-}(x,\bfr)&=& \int \frac{d^2 \bfk}{(2 \pi)^2} e^{i \bfk \cdot \bfr}
\psi^{+}_{-q}(x, \bfk)\nonumber\\
&=& - i \frac{N^{2}_{q}}{2 \pi} \  \frac{4 \pi}{\kappa}
\sqrt{\frac{\log(1/x)}{1-x}} \frac{x^{a_{q}^{2}} (1-x)^{b_{q}^{2}}}{M x} \frac{(r^1+ir^2)
e^{\frac{- r^2}{4g}}}{4 g^2},\label{Co-WF} \nonumber\\
\tilde{\psi}^{-}_{q+}(x,\bfr)&=& \int \frac{d^2 \bfk}{(2 \pi)^2} e^{i \bfk \cdot \bfr}
\psi^{-}_{+q}(x, \bfk)\nonumber\\
&=&  i \frac{N^{2}_{q}}{2 \pi} \  \frac{4 \pi}{\kappa}
\sqrt{\frac{\log(1/x)}{1-x}} \frac{x^{a_{q}^{2}} (1-x)^{b_{q}^{2}}}{M x} \frac{(r^1-ir^2)
e^{\frac{- r^2}{4g}}}{4 g^2}, \nonumber\\
\tilde{\psi}^{-}_{q-}(x,\bfr)&=& \int \frac{d^2 \bfk}{(2 \pi)^2} e^{i \bfk \cdot \bfr}
\psi^{-}_{-q}(x, \bfk) \nonumber\\
&=&  \frac{N^{1}_{q}}{2 \pi} \  \frac{4 \pi}{\kappa} \sqrt{\frac{\log(1/x)}{1-x}}
\frac{x^{a_{q}^{2}} (1-x)^{b_{q}^{2}}}{M x} \frac{e^{\frac{- r^2}{4g}}}{2 g},
\ee
where $g= \frac{\log(1/x)}{2 \kappa^2 (1-x)^2}$. Substituting Eq.(\ref{Co-WF}) in
Eq.(\ref{defi}), we can evaluate the charge distributions for a particular flavor in transverse coordinate
space and is given as
\be
P_{f}^q(r)=\int dx
\frac{\log(1/x)}{(1-x)}\bigg[(N^1_q)^2x^{2a^1_q}(1-x)^{2b^1_q}+(N^2_q)^2\frac{x^{2a^2_q}(1-x)^{
2b^2_q}}{x^2M_n^2}\frac{r^2}{4g^2}\bigg]\frac{1}{\kappa^2g^2}e^{\frac{- r^2}{4g}}.
\ee
As for the case of decomposition for Dirac and Pauli form factors in Eq.(\ref{N_ff}), using charge
and isospin symmetry, we can write the charge
distributions for the nucleons \cite{hwang,CM3} as follows
\be
P^{p}&=& e_u  P_{f}^{u}+ e_d  P_{f}^{d}=\frac{4}{3}P^u-\frac{1}{3}P^d, \nonumber\\
P^{n}&=& e_u  P_{f}^{d}+ e_d  P_{f}^{u}=\frac{2}{3}P^d-\frac{2}{3}P^u,
\ee
where $P_{f}^q(r)$ and $P^q$ are the charge distributions for a particular flavor present in the nucleon and the individual quark
respectively. Thus, the individual quark distributions are related to the charge distribution of a particular flavor present in the proton is
\be
P^{u}&=& \frac{P_{f}^{u}}{2}, \nonumber\\
P^{d}&=& P_{f}^{d}.
\ee

\section{Charge densities in impact-parameter space}\label{charge-density}
According to the standard interpretation \cite{miller07,vande,weiss,CM3,selyugin}, the charge
density in the impact-parameter space can be identified with the two-dimensional
Fourier transform of the Dirac form factor in the light-cone frame with $q^+=q^0+q^3=0$
\be
\rho_{ch}(b)
&=&\int \frac{d^2\bfq}{(2\pi)^2}F_1(q^2)e^{i\bfq\cdot\bfb},
\ee
where the impact-parameter, $b=|\bfb|$. Using the Dirac form factor in term of overlaps of LFWFs
defined in Eq.(\ref{Dirac_FF}), we can write the charge density for a particular flavor in the nucleon as
\be\label{ch_wf}
\rho^q_{fch}(b)
&=&\int \frac{d^2\bfq}{(2\pi)^2}e^{i\bfq\cdot\bfb}\frac{1}{4\pi}\int_0^1dx \int
\frac{d^2\bfk}{(2\pi)^2}~\bigg[\psi_{+q}^{+*}(x,\bfk')\psi_{+q}^+(x,\bfk) \nonumber\\
&&+\psi_{-q}^{+*}(x,\bfk')\psi_{-q}^+(x,\bfk)\bigg].
\ee
Now, using the elementary theorem of convolutions and Fourier
transform (see the appendix \ref{appen}), the above Eq.(\ref{ch_wf}) can be re-written as
\cite{hwang}
\be\label{rel}
\rho^q_{fch}(b)\label{rel_charge}
&=&\frac{1}{4\pi}\int_0^1dx
\frac{1}{(1-x)^2}\bigg[\tilde{\psi}^{+*}_{+q}\Big(x,\frac{\bfb}{x-1}\Big)\tilde{\psi}^{+}_{+q}
\Big(x,\frac{\bfb}{x-1}\Big) \nonumber\\
&&+
\tilde{\psi}^{+*}_{-q}\Big(x,\frac{\bfb}{x-1}\Big)\tilde{\psi}^{+}_{-q}\Big(x,\frac{\bfb}{x-1}
\Big)\bigg],\nonumber\\
&=&\frac{1}{4\pi}\int_0^1dx \frac{1}{(1-x)^2}P^q_{f}\Big(x,\frac{\bfb}{x-1}\Big).
\ee
This relation shows that the coordinate space density is not the same as impact parameter space density. Though both $r$ and $|b_\perp|$ are conjugate to momenta, there is still a distinction between these two. $r$ is conjugate to the momentum $k$ whereas $b_\perp$ is conjugate to the momentum transferred $\Delta_\perp$.  So,  interpreting the density in impact parameter space as the coordinate space density is not correct.
Similar calculations have been carried out in Ref. \cite{hwang} where they have calculated the charge distribution in
transverse coordinate space, however the factor $\frac{1}{4\pi}$ is missing in the
relation between charge distributions in
impact-parameter space and the transverse coordinate space. Now, the
charge densities for nucleons can be written in term of flavor densities  \cite{CM3,CM4} as follows
\be
 \rho_{ch}^p&=& e_u \rho_{fch}^u+e_d \rho_{fch}^d,\nonumber\\\label{ch_mag1}
 \rho_{ch}^n&=& e_u \rho_{fch}^d+e_d \rho_{fch}^u.
 \ee
Due to the charge and isospin symmetry, the $u$ and  $d$ quark densities in the proton are same as
the $d$ and $u$ quark densities in the neutron \cite{miller07,CM3}. Under the charge and isospin
symmetry, one can write
\be
 \rho_{ch}^u(b)&=&  \rho_{ch}^p+ \frac{\rho_{ch}^n}{2}=\frac{
\rho_{fch}^u}{2},\nonumber\\\label{ch_mag2}
 \rho_{ch}^d(b)&=&  \rho_{ch}^p+2 \rho_{ch}^n= \rho_{fch}^d,
 \ee
where $\rho_{ch}^q(b)$ is the charge density of each quark and $\rho^q_{fch}$ is the charge
density for a particular flavor.

\begin{figure}[htbp]
\begin{minipage}[c]{0.98\textwidth}
{(a)}\includegraphics[width=7.5cm,clip]{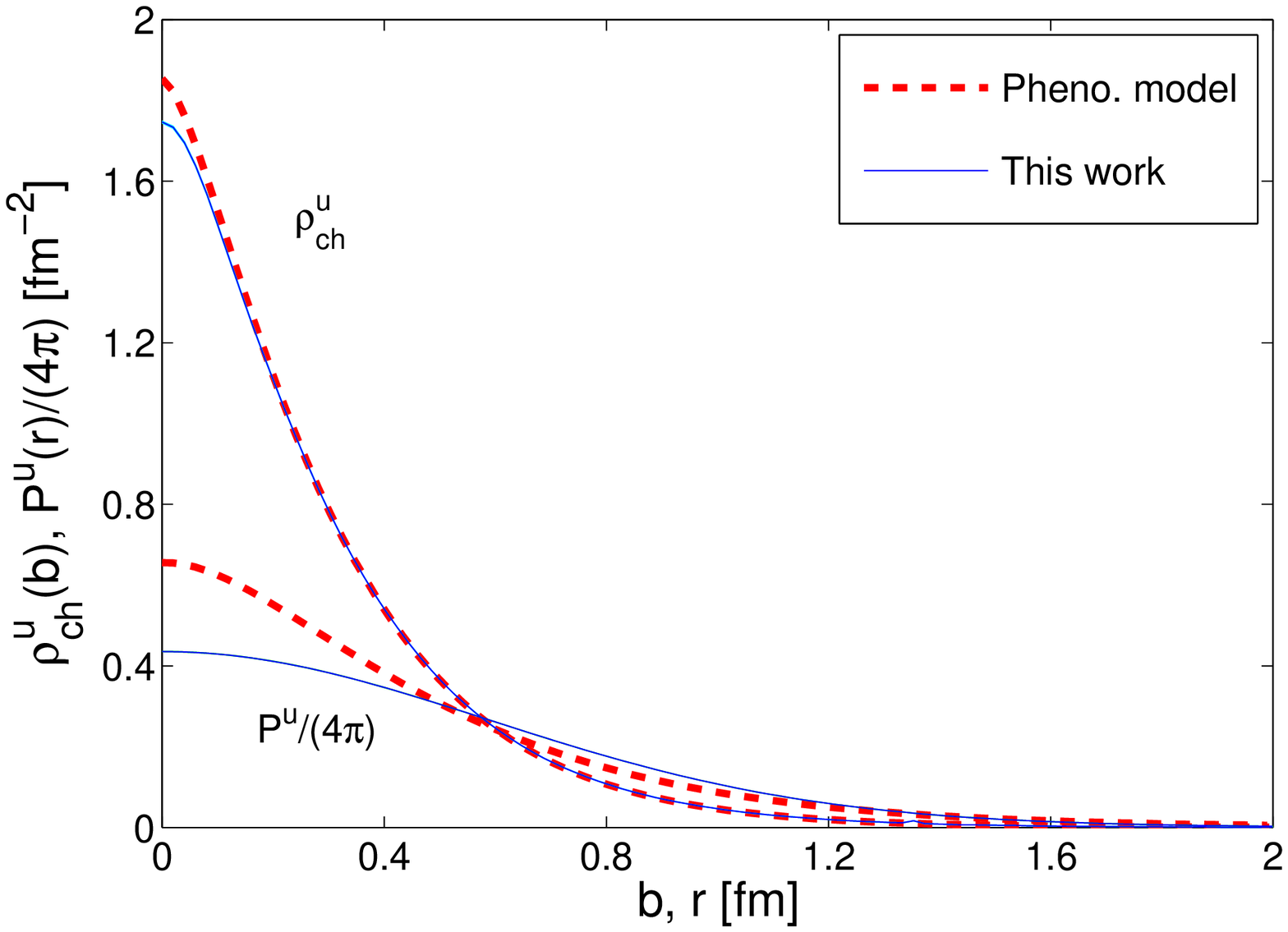}
{(b)}\includegraphics[width=7.5cm,clip]{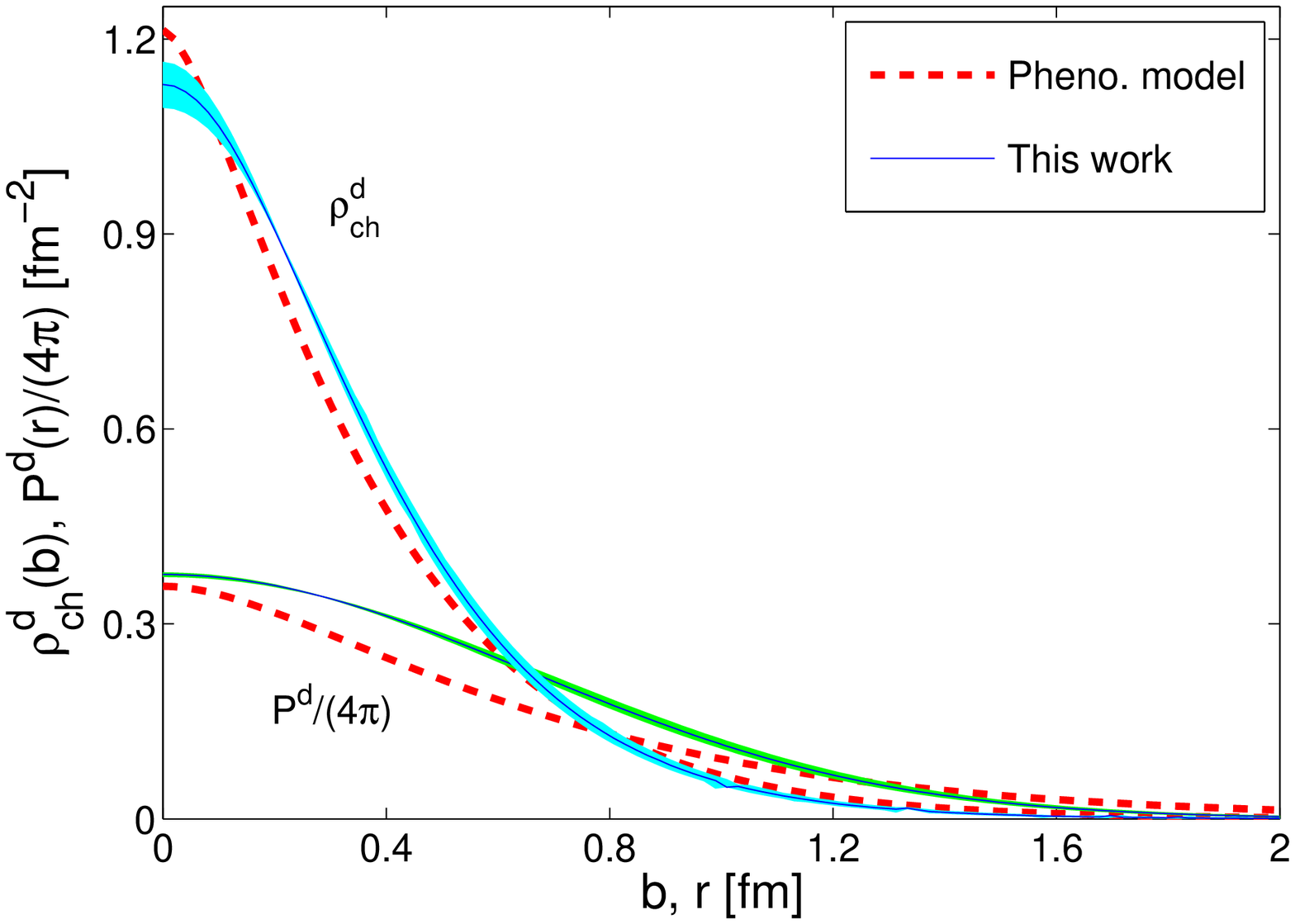}
\end{minipage}
\caption{\label{plot1}(colour online) Charge distribution $P^q(\bfr)$ and $\rho^q_{ch}(\bfb)$ in the transverse coordinate and impact-parameter spaces respectively for the (a) $u$ quarks and (b) $d$ quarks, where $r=|\bfr|$ and $b=|\bfb|$. The blue solid lines represent the quark-diquark model in AdS/QCD (this work) and the red dashed lines represent the phenomenological model \cite{hwang}. The bands in the plots represent the errors in the model predictions. For $u$ quark the bands are very small.}
\end{figure}
\begin{figure}[htbp]
\begin{minipage}[c]{0.98\textwidth}
{(a)}\includegraphics[width=7.5cm,clip]{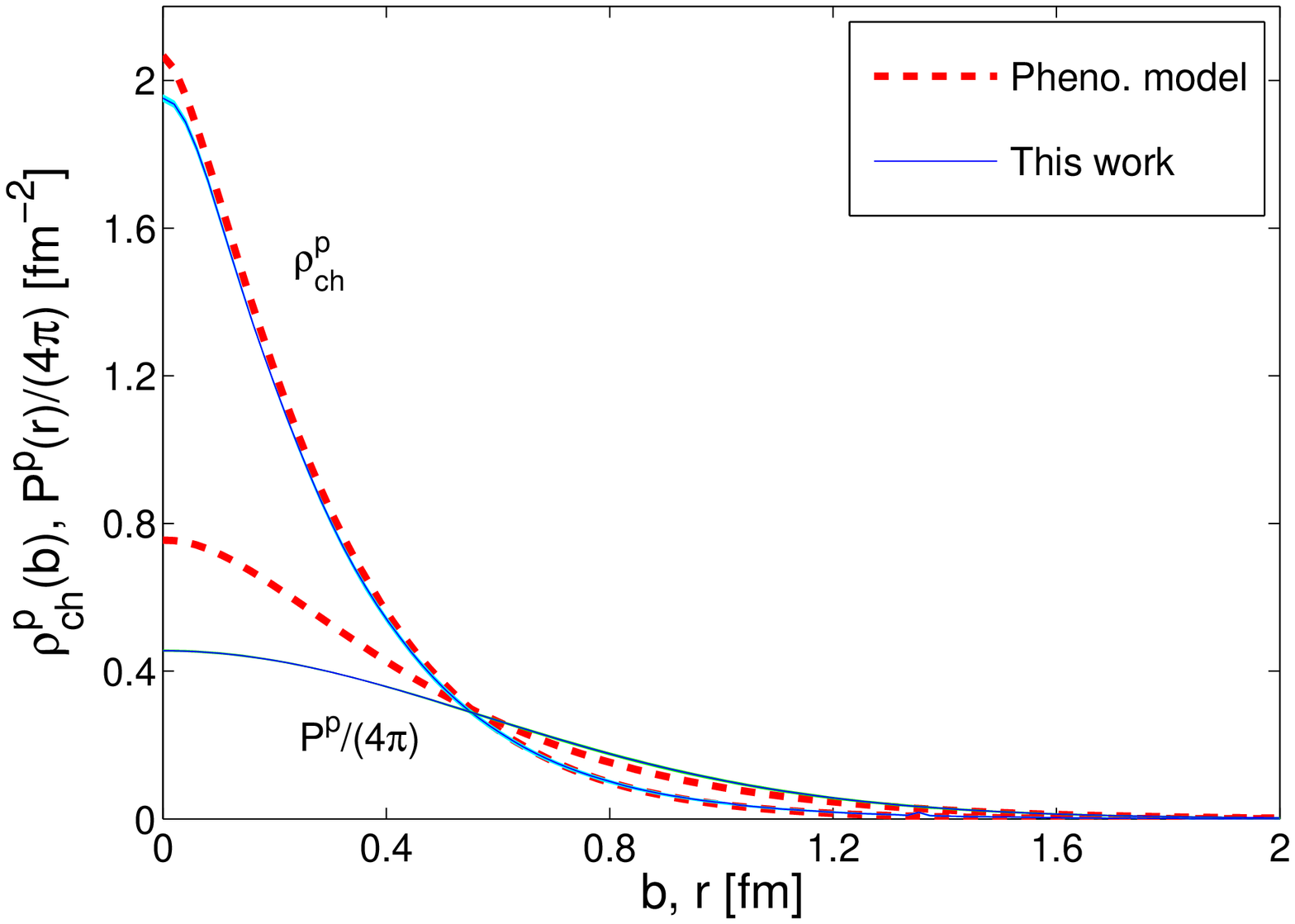}
{(b)}\includegraphics[width=7.5cm,clip]{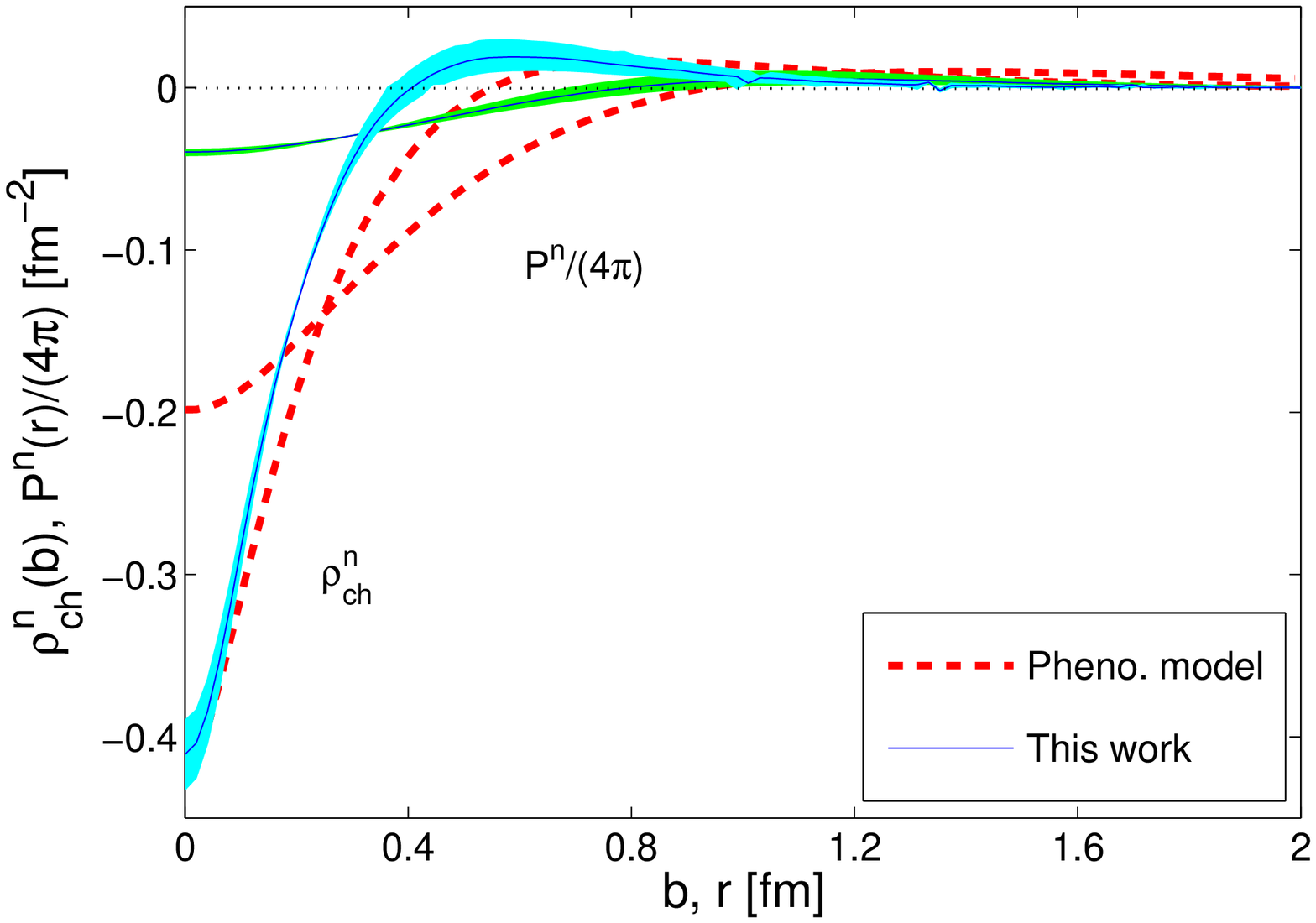}
\end{minipage}
\caption{\label{plot2}(colour online) Charge distribution $P^{p,n}(\bfr)$ and $\rho^{p,n}_{ch}(\bfb)$ in the transverse coordinate and
impact-parameter spaces respectively for the (a) proton and  (b)  neutron. The blue solid lines represent the quark-diquark model in AdS/QCD (this work) and the red dashed lines represent the phenomenological model \cite{hwang}.The bands in the plots represent the errors in the model predictions.}
\end{figure}
\begin{figure}[htbp]
\begin{minipage}[c]{0.98\textwidth}
{(a)}\includegraphics[width=7.5cm,clip]{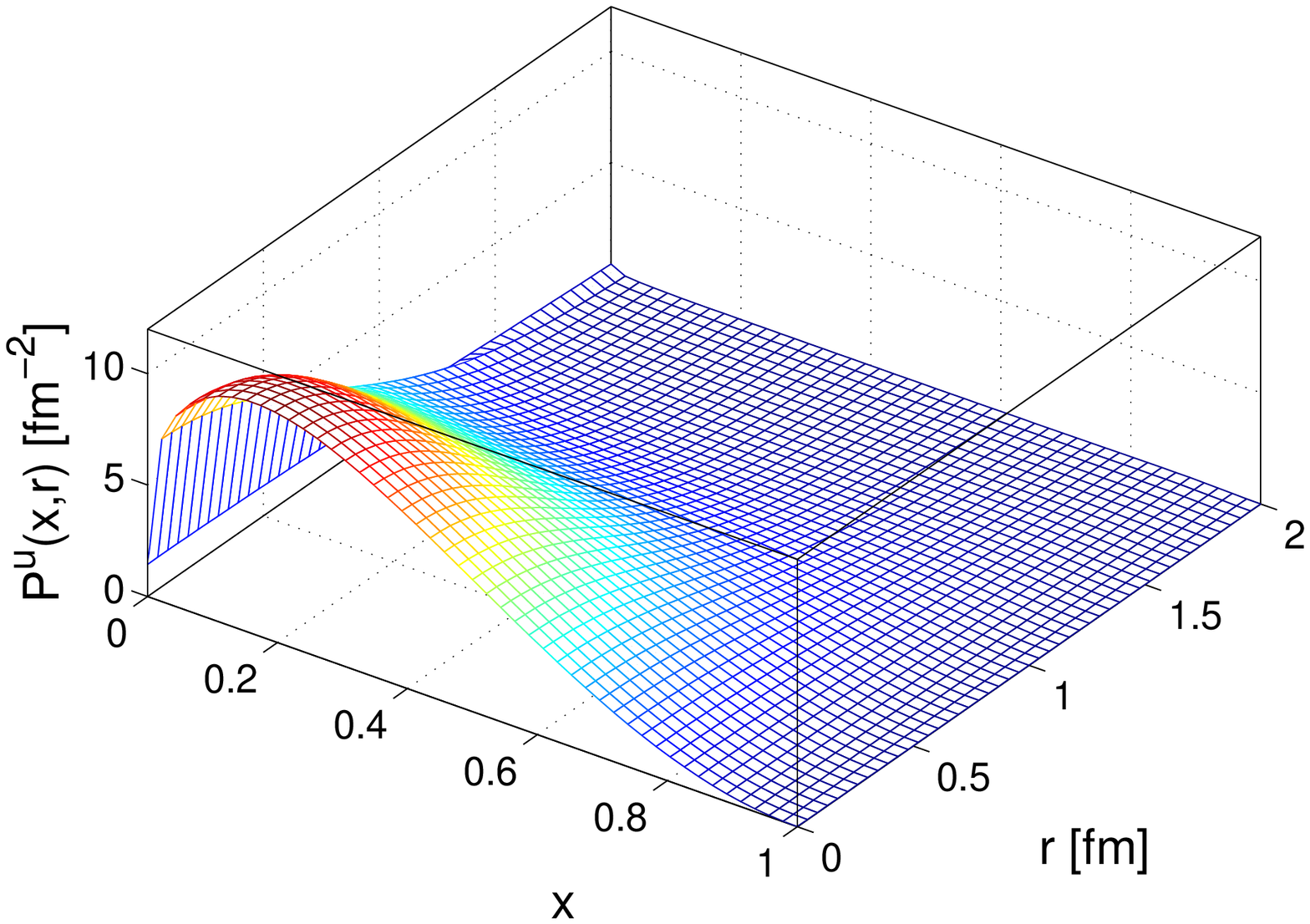}
{(b)}\includegraphics[width=7.5cm,clip]{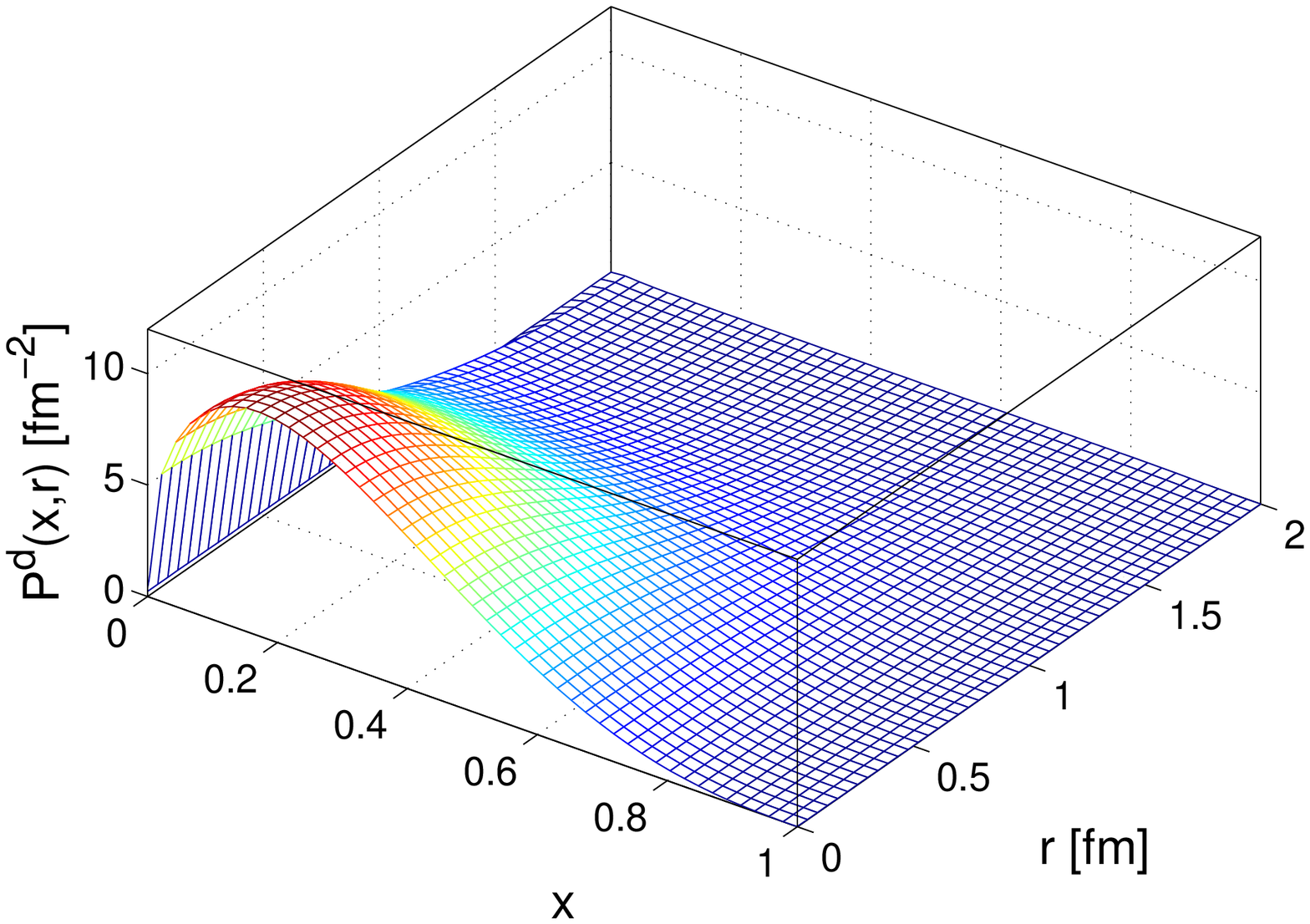}
\end{minipage}
\caption{\label{plot3}(colour online) Charge distribution (a) $P^u(x,\bfr)$ and (b)
$P^d(x,\bfr)$ in the transverse coordinate space as a function of $\bfr $
and $x$.}
\end{figure}
\begin{figure}[htbp]
\begin{minipage}[c]{0.98\textwidth}
{(a)}\includegraphics[width=7.5cm,clip]{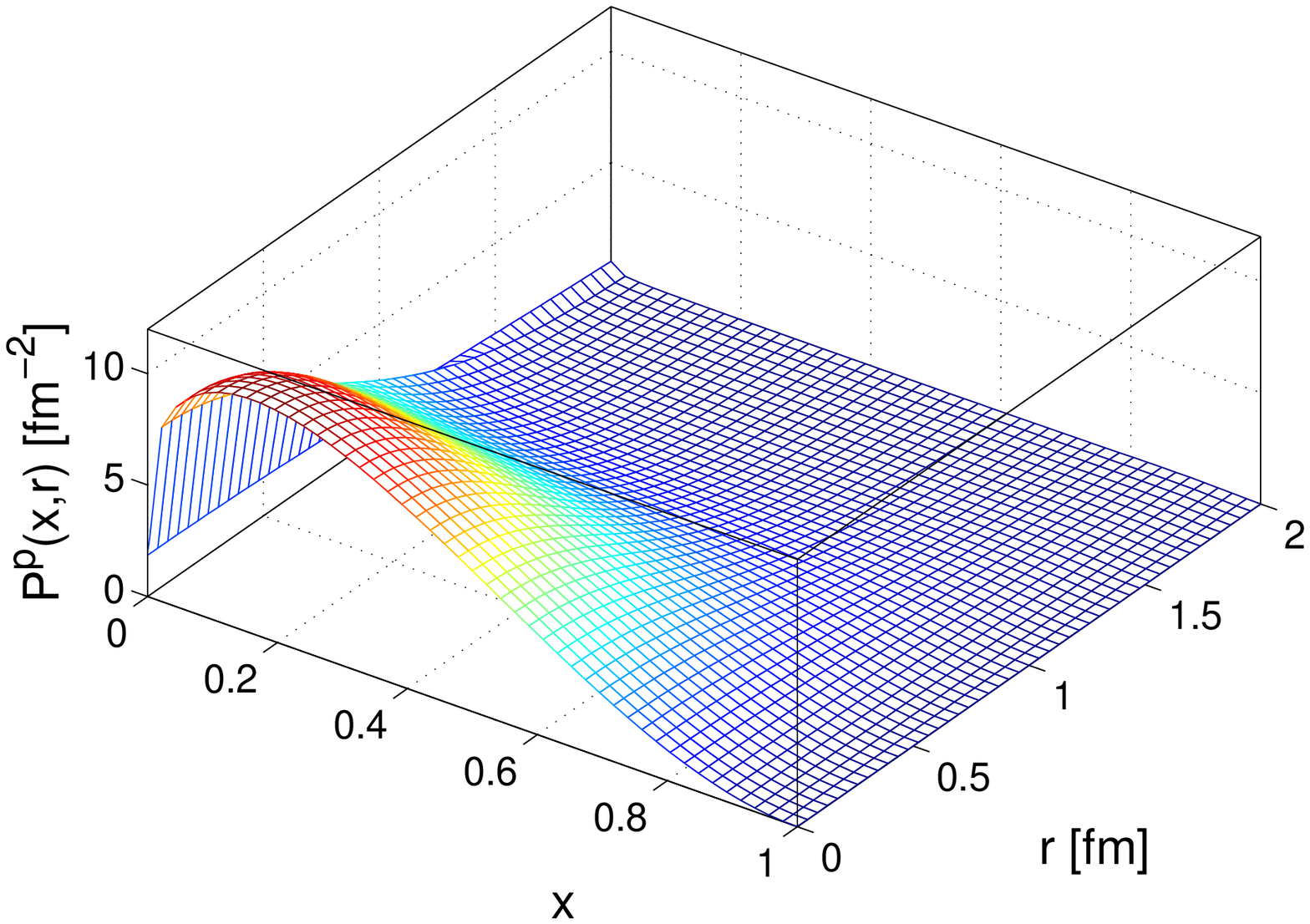}
{(b)}\includegraphics[width=7.5cm,clip]{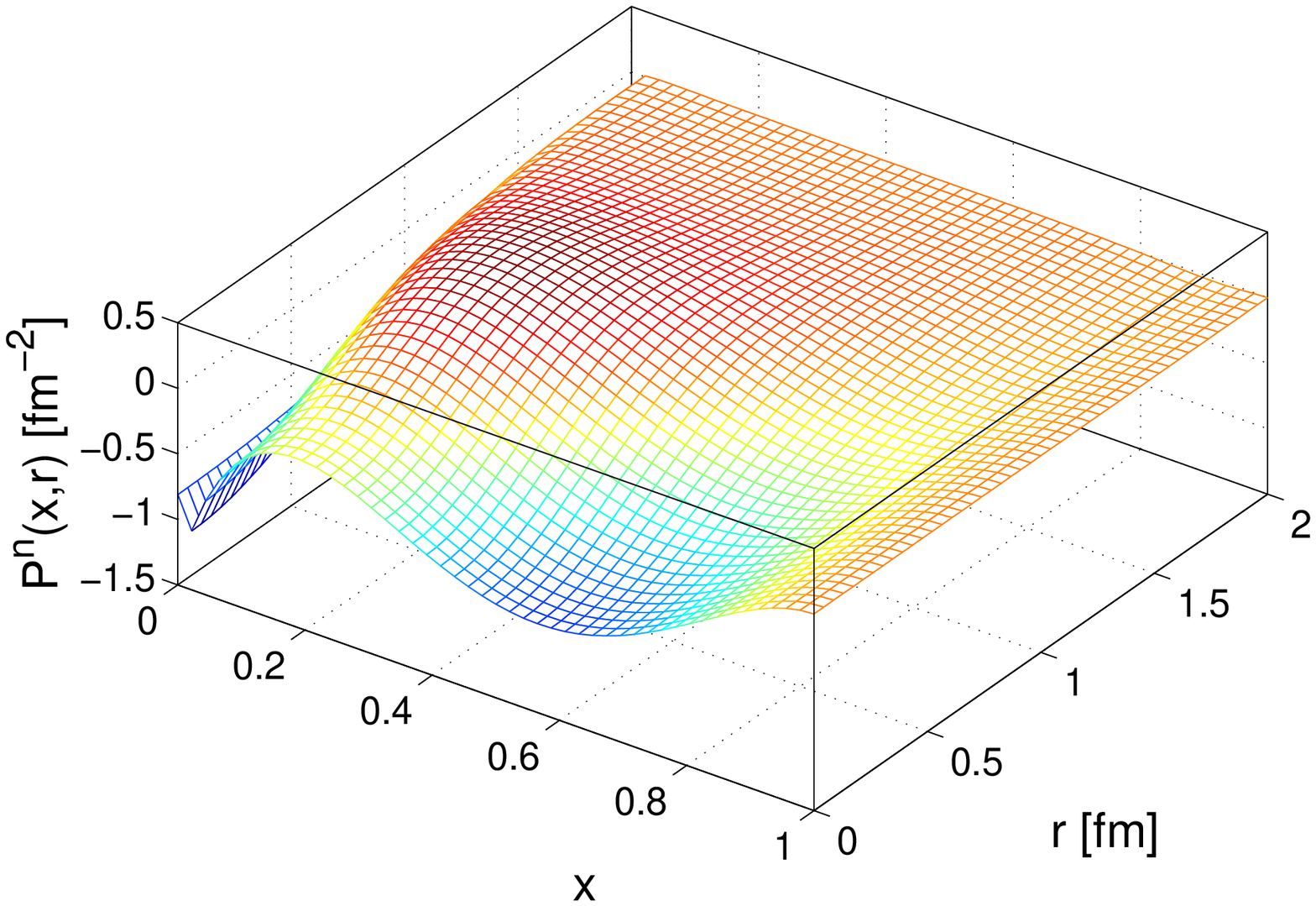}
\end{minipage}
\caption{\label{plot4}(colour online) Charge distribution (a) $P^p(x,\bfr)$ and (b) $P^n(x,\bfr)$ in the
transverse coordinate space as a function of $\bfr $
and $x$.}
\end{figure}
In Fig.\ref{plot1}, we show the charge distributions for $u$ and $d$
quarks in the transverse coordinate space. The charge distributions for $u$ and $d$ in
impact-parameter space are also shown in the same plot.
With $2\sigma$ error in the model parameters, we estimate the errors in the model predictions. The bands in the plots show the error in our model.
For $u$ quark the  errors come out to be very small with $2\sigma$ uncertainty in the model parameters.
We observe that the magnitude of charge distributions in impact-parameter is larger than the distributions in transverse coordinate space. The charge distributions for
proton and neutron in both transverse coordinate and impact-parameter spaces are
plotted in Fig.\ref{plot2}.
Since these quantities are not directly measured in experiments, actual experimental data are not available. Thus, we compare the distributions for both quarks and nucleon with the results of another phenomenological model \cite{hwang}.
We observe that  the quark-diquark model in AdS/QCD is in good agreement with the phenomenological model for the charge distributions in impact parameter plane.
Though, one can see from Fig.\ref{plot2}(b), that the two models do not agree for the charge distributions for neutron and
at low $\bfr$, the phenomenological model gives higher values of  $P(r)$ (except $P^d$), the overall behaviors of the distributions in both the models are qualitatively the same.
One can notice that the width of the distributions in coordinate
space is larger than that in impact-parameter space. It can be understood from the relation
between these two distributions shown in Eq.(\ref{rel}) where $\frac{\bfb}{x-1}$ appears in place
of $\bfr$ in the argument of $P(x,\bfr)$. We also observe that the proton (neutron) charge
density and distribution has a long range positively (negatively) charged component. The charge distribution for neutron is observed to be mostly negative though it also shows a slightly positive distribution whereas for proton it remains positive. This shows that charge
distribution for neutron has a negatively charged core surrounded by a positively charged shell. For proton,
we notice that, the density is peaked at low values of $b$ and  has a long positive tail. Similar behavior of the
charge distributions has been observed in other phenomenological models \cite{hwang,kumar}.

In order to get more information about the distributions
$P(x,\bfr)$, we plot the distributions as a function of $x$ and $\bfr$. In Fig.\ref{plot3}, we
show the distributions $P(x,\bfr)$ for individual $u$ and $d$ quarks as a function of $x$ and $\bfr$. Similar plots for proton and
neutron are shown in Fig.\ref{plot4}.
We observe that all the distributions are peaked at lower $x$ and with increasing the value of
$x$, the magnitudes of the charge distributions
decrease. However, the magnitude of the distributions for $u$ and $d$ quarks are more or less
same.

\section{longitudinal momentum distribution: coordinate space}\label{longi_coordinate}
For a spin-half composite system, similar to the electromagnetic form factors, the
gravitational form factors $A(Q^2)$ and $B(Q^2)$ can be obtained from the helicity conserving and
helicity-flip matrix elements of the $T^{++}$ tensor current. $A(Q^2)$ and $B(Q^2)$ are analogous
to $F_1(Q^2)$ (Dirac) and $F_2(Q^2)$ (Pauli) form factors for the $J^+$ vector current. The
helicity conserved form factor $A(Q^2)$ allows one to measure the momentum fractions carried by
each constituent of a hadron. The gravitational form factors in the light-front quark-diquark
model can be obtained in terms of the overlap of the wave functions as \cite{chakrabarti,BHMI}
\be\label{GFF}
A^q(Q^2) &=& \int_0^1dx \int
\frac{d^2\bfk}{16\pi^3}~x\Big[\psi_{+q}^{+*}(x,\bfk')\psi_{+q}^+(x,\bfk)
+\psi_{-q}^{+*}(x,\bfk')\psi_{-q}^+(x,\bfk)\Big],\label{A_FF}\\
B^q(Q^2) &=& -\frac{2M_n}{q^1-iq^2}\int_0^1dx \int \frac{d^2\bfk}{16\pi^3}~x
\Big[\psi_{+q}^{+*}(x,\bfk')\psi_{+q}^-(x,\bfk)\nonumber\\
&+&\psi_{-q}^{+*}(x,\bfk')\psi_{-q}^-(x,\bfk)\Big]. \ee
The $++$ component of the energy-momentum tensor
$(T^{\mu \nu})$ provides the longitudinal momentum
\begin{equation}
P^+= \int dx^- d^2 x^\perp T^{++}.
\end{equation}
Using the LFWFs in transverse coordinate space we
can now define the longitudinal momentum distribution in the transverse coordinate space for
a struck flavor $q$ (fermionic)
\be \label{defi_longi}
P_{Lq(flavor)}^f(r)&=& \int dx P_L^q(x,r) = \int dx
~x\left[\tilde{\psi}^{+*}_{+q}(x,\bfr)\tilde{\psi}^{+}_{+q}(x,\bfr) +
\tilde{\psi}^{+*}_{-q}(x,\bfr)\tilde{\psi}^{+}_{-q}(x,\bfr)\right]\nonumber\\
&=& \int dx~x~ \frac{\log(1/x)}{(1-x)}\bigg[(N^1_q)^2x^{2a^1_q}(1-x)^{2b^1_q}+(N^2_q)^2\frac{x^{2a^2_q}(1-x)^{2b^2_q}}{x^2M_n^2}\frac{r^2}{4g^2}\bigg]\frac{1}{\kappa^2g^2}e^
{\frac{- r^2}{4g}}, \nonumber\\
\ee
and for the diquark (bosonic)
\be \label{longi_b}
P_{Lq(flavor)}^b(r)&=& \int dx P_L^b(x,r) = \int dx
~(1-x)\left[\tilde{\psi}^{+*}_{+q}(x,\bfr)\tilde{\psi}^{+}_{+q}(x,\bfr) +
\tilde{\psi}^{+*}_{-q}(x,\bfr)\tilde{\psi}^{+}_{-q}(x,\bfr)\right]\nonumber\\
&=& \int dx~(1-x)~ \frac{\log(1/x)}{(1-x)}\bigg[(N^1_q)^2x^{2a^1_q}(1-x)^{2b^1_q}+(N^2_q)^2\frac{x^{2a^2_q}(1-x)^{2b^2_q}}{x^2M_n^2}\frac{r^2}{4g^2}\bigg]\frac{1}{\kappa^2g^2}e^
{\frac{- r^2}{4g}}. \nonumber\\
\ee
Thus, the longitudinal momentum distributions for the proton in transverse coordinate space for the
struck $u$ and $d$ quarks are given as
\be
P_{Lu}^p(r)&=&\frac{1}{2} [P_{Lu(flavor)}^f+P_{Lu(flavor)}^b]=P_{Lu}^f+P_{Lu}^b,\nonumber\\
P_{Ld}^p(r)&=&[P_{Ld(flavor)}^f+P_{Ld(flavor)}^b]=P_{Ld}^f+P_{Ld}^b.
\ee
For $P_{Lu}^p$, the $\frac{1}{2}$ factor appears due to two possibilities of struck quark being a $u$ quark
(two valence $u$ quark in proton).
\begin{figure}[htbp]
\begin{minipage}[c]{0.98\textwidth}
{(a)}\includegraphics[width=7.5cm,clip]{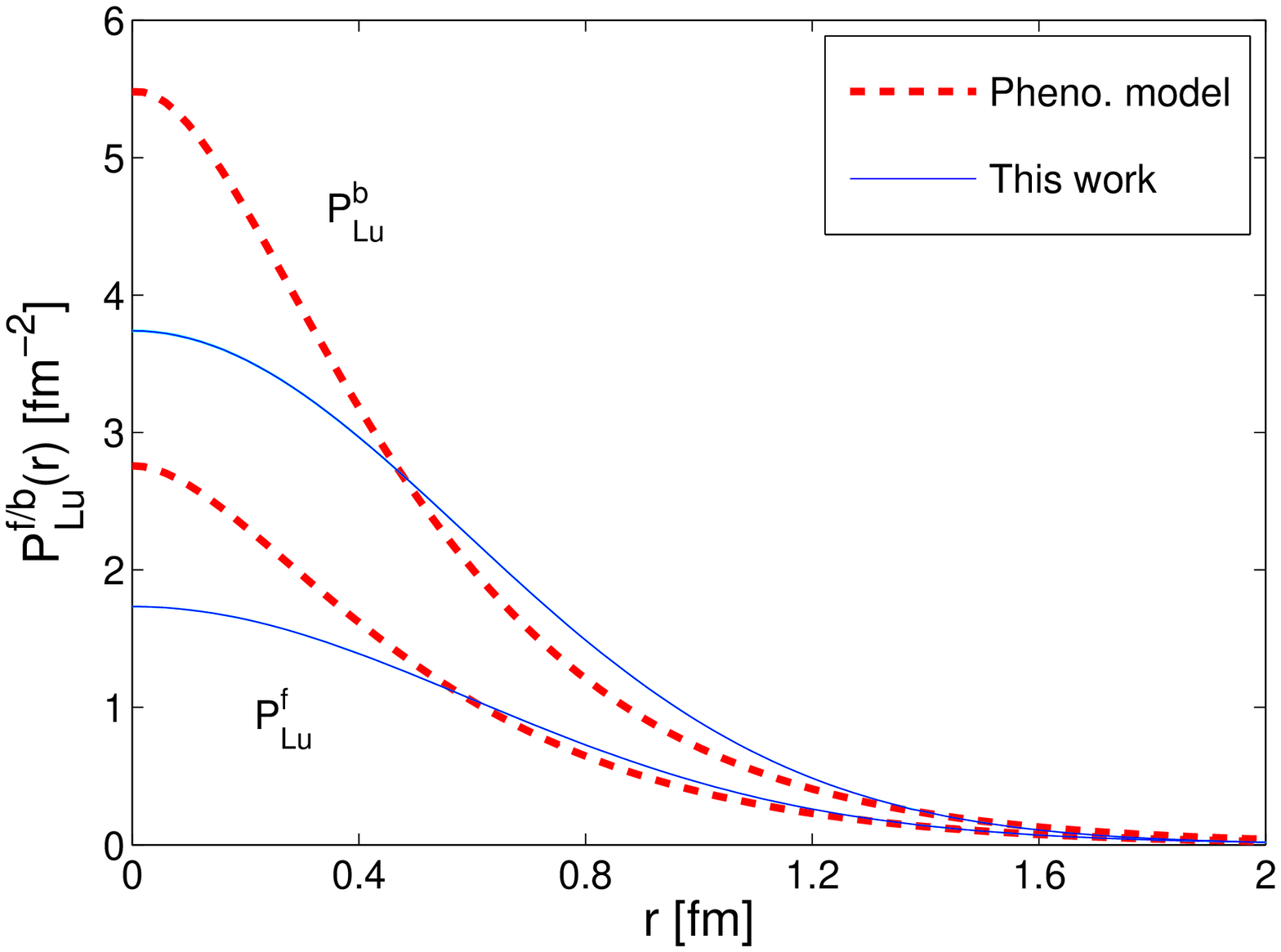}
{(b)}\includegraphics[width=7.5cm,clip]{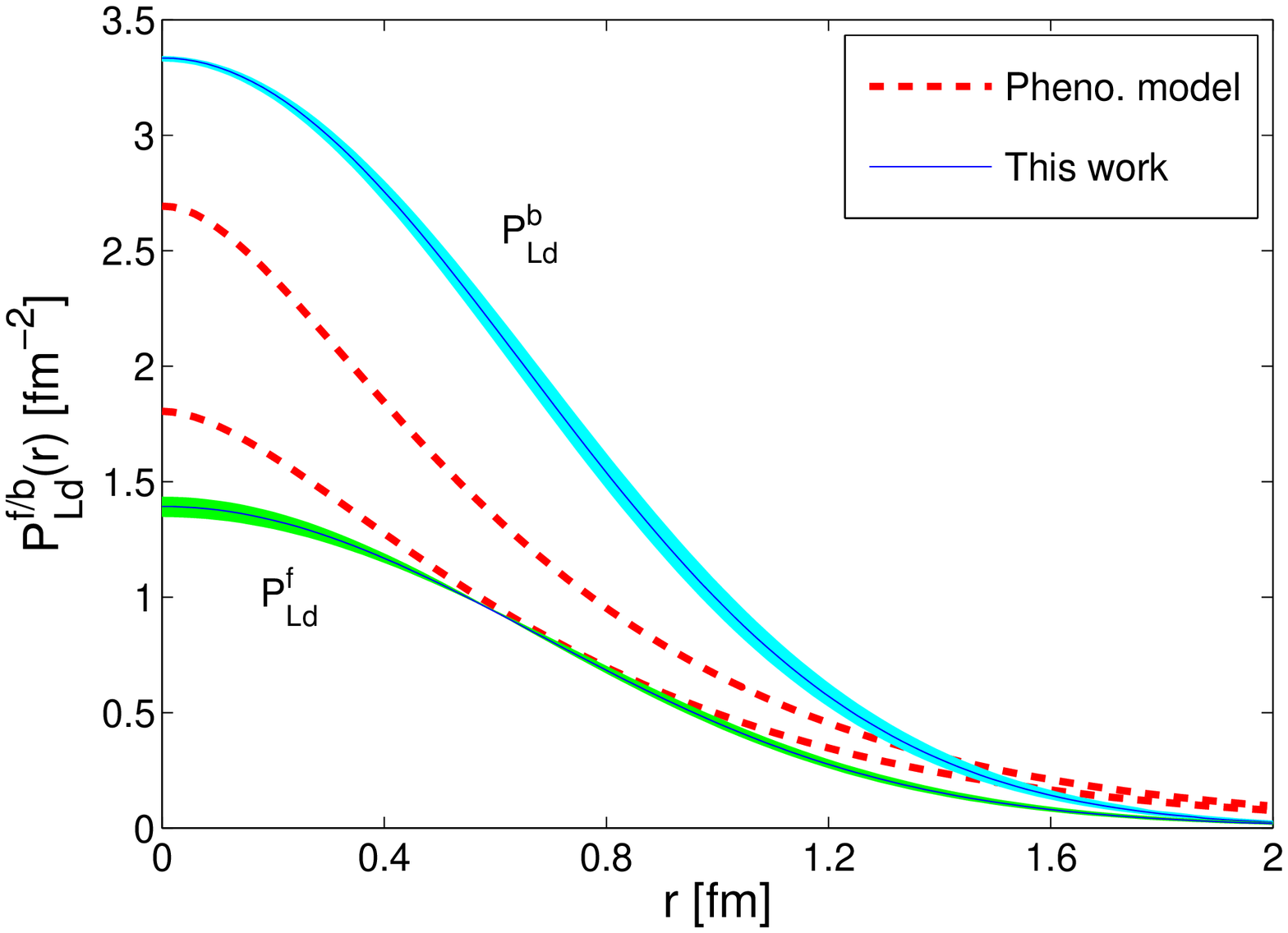}
\end{minipage}
\begin{minipage}[c]{0.98\textwidth}
\begin{center}
{(c)}\includegraphics[width=7.5cm,clip]{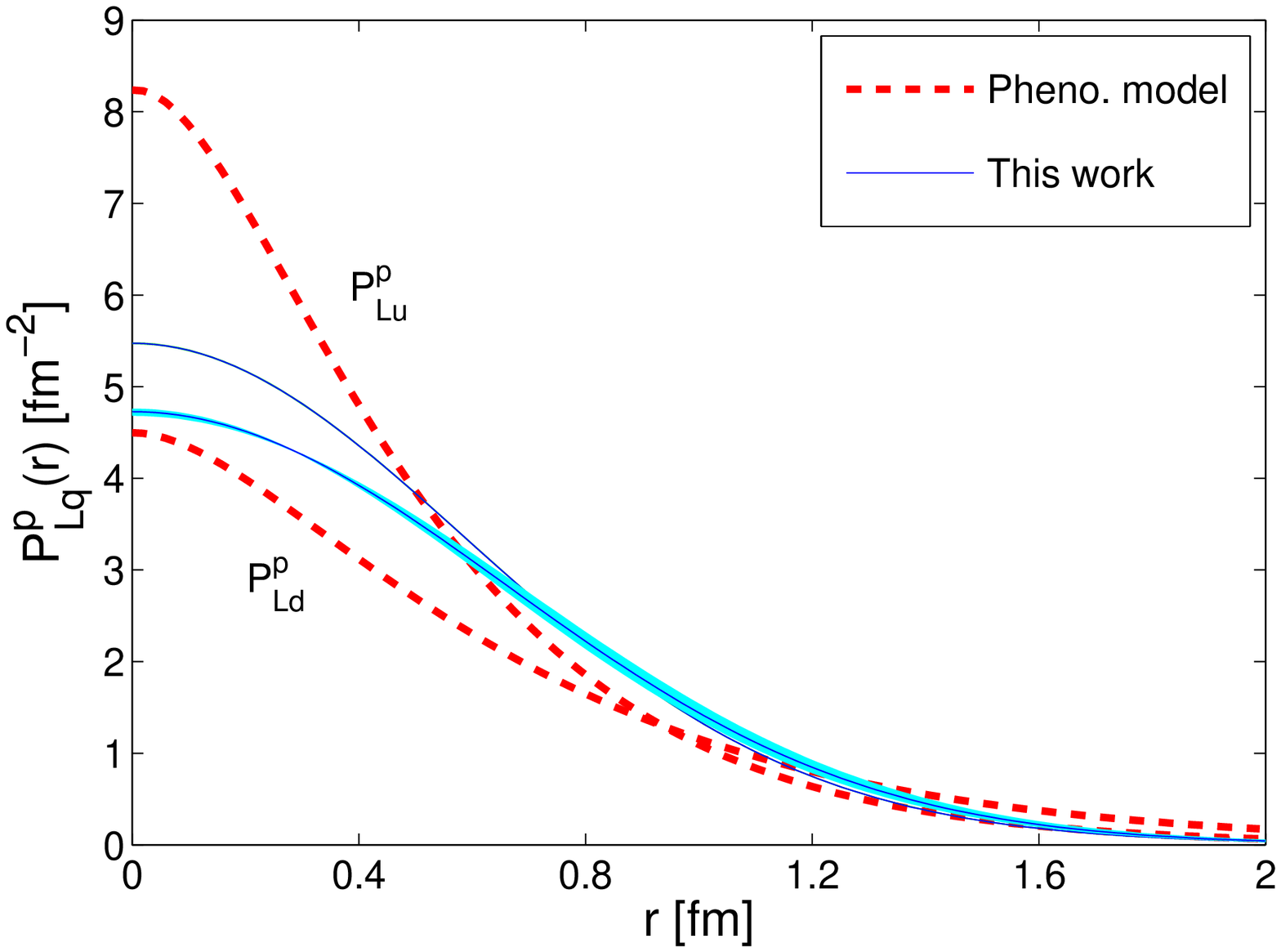}
\end{center}
\end{minipage}
\caption{\label{plot5}(colour online) Longitudinal momentum distributions (a) $P^{f/b}_L$
for struck $u$ quark, (b) $P^{f/b}_L$  for struck $d$ quark $d$ and (c)  $P^{p}_L$ for proton when struck
quarks are $u$ and $d$. The blue solid lines represent the quark-diquark model in AdS/QCD (this work) and the red dashed lines represent the phenomenological model \cite{hwang}. The error bands shown in the plots are estimated from the errors in the model parameters.}
\end{figure}
\begin{figure}[htbp]
\begin{minipage}[c]{0.98\textwidth}
{(a)}\includegraphics[width=7.5cm,clip]{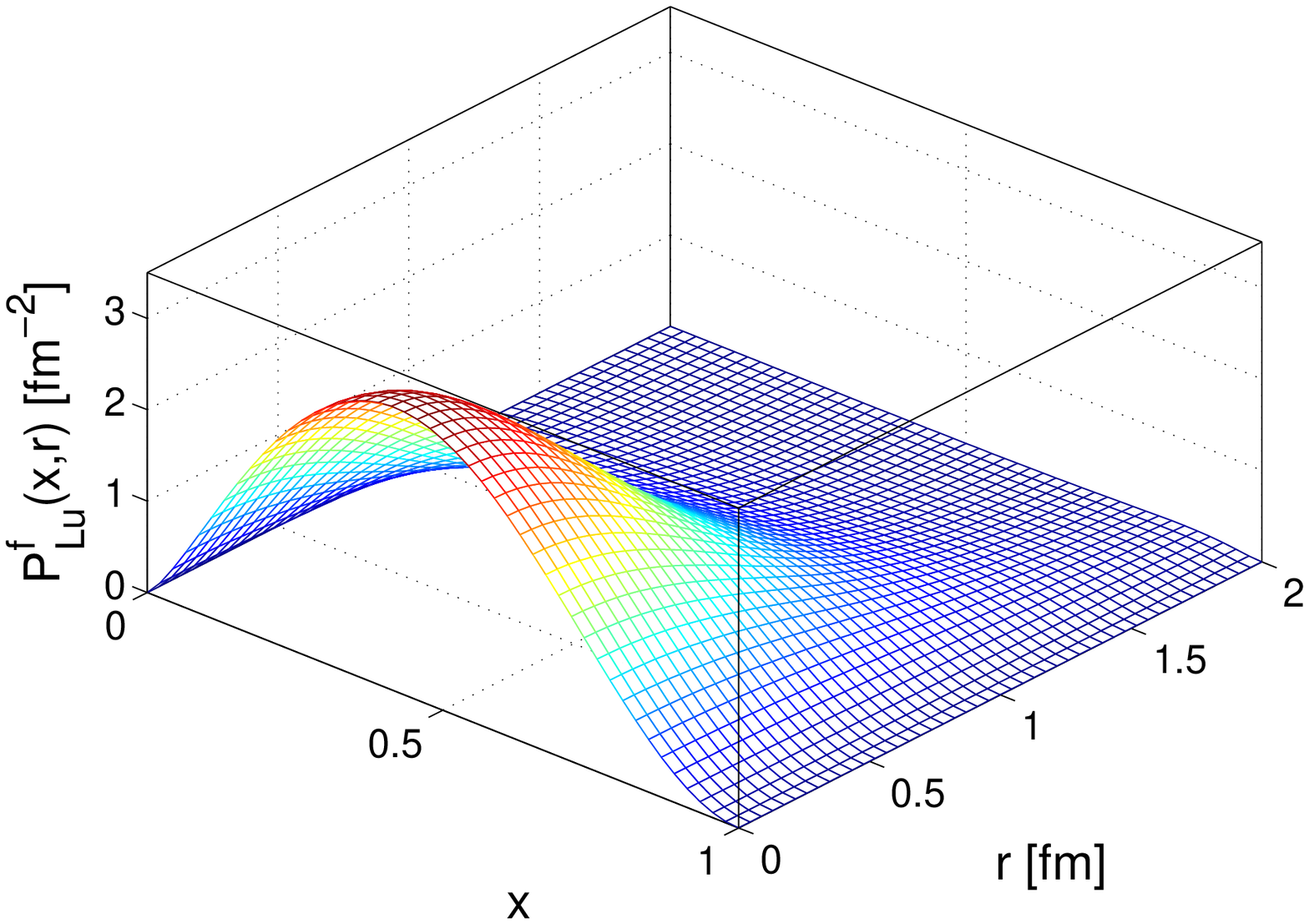}
{(b)}\includegraphics[width=7.5cm,clip]{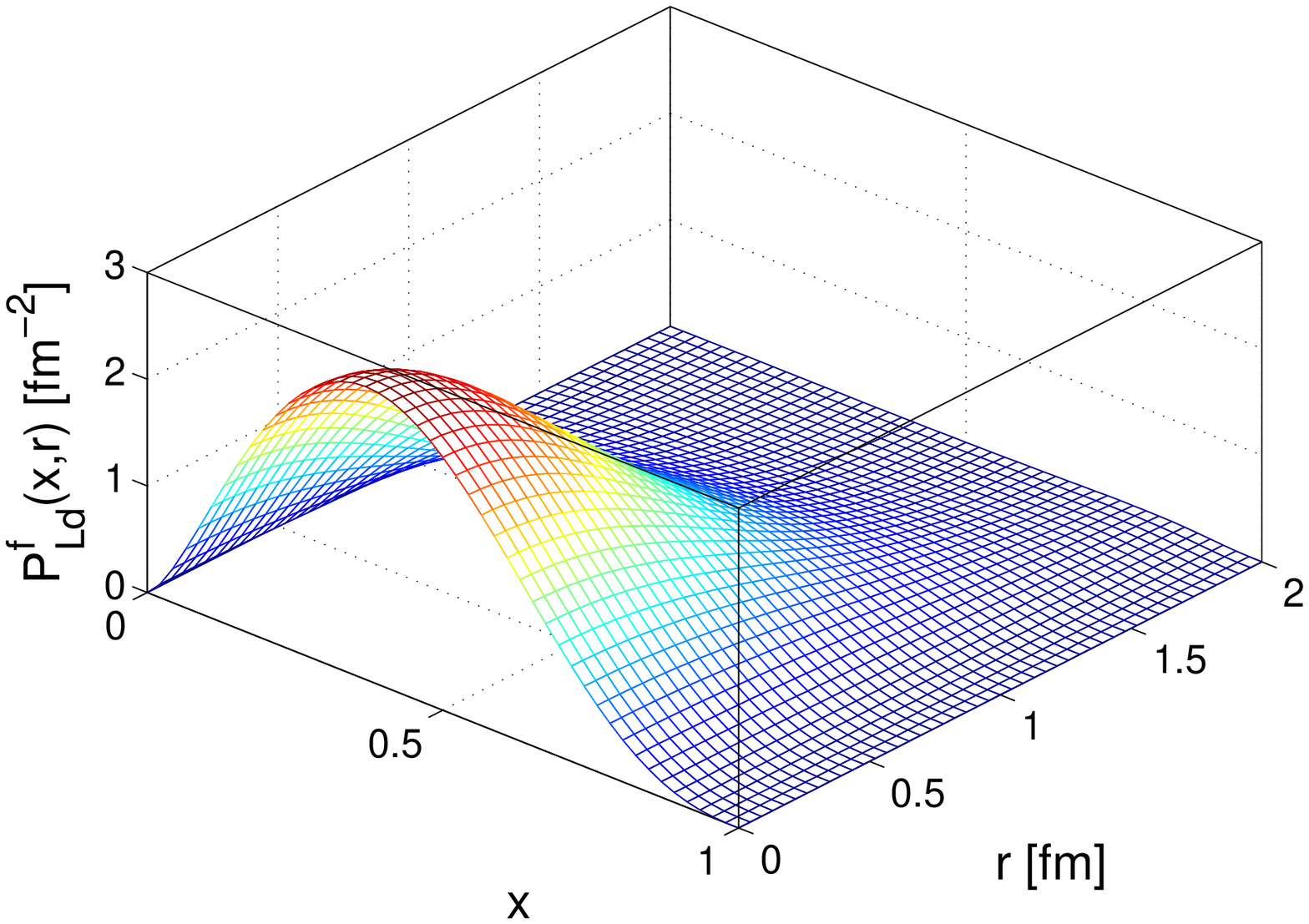}
\end{minipage}
\begin{minipage}[c]{0.98\textwidth}
{(c)}\includegraphics[width=7.5cm,clip]{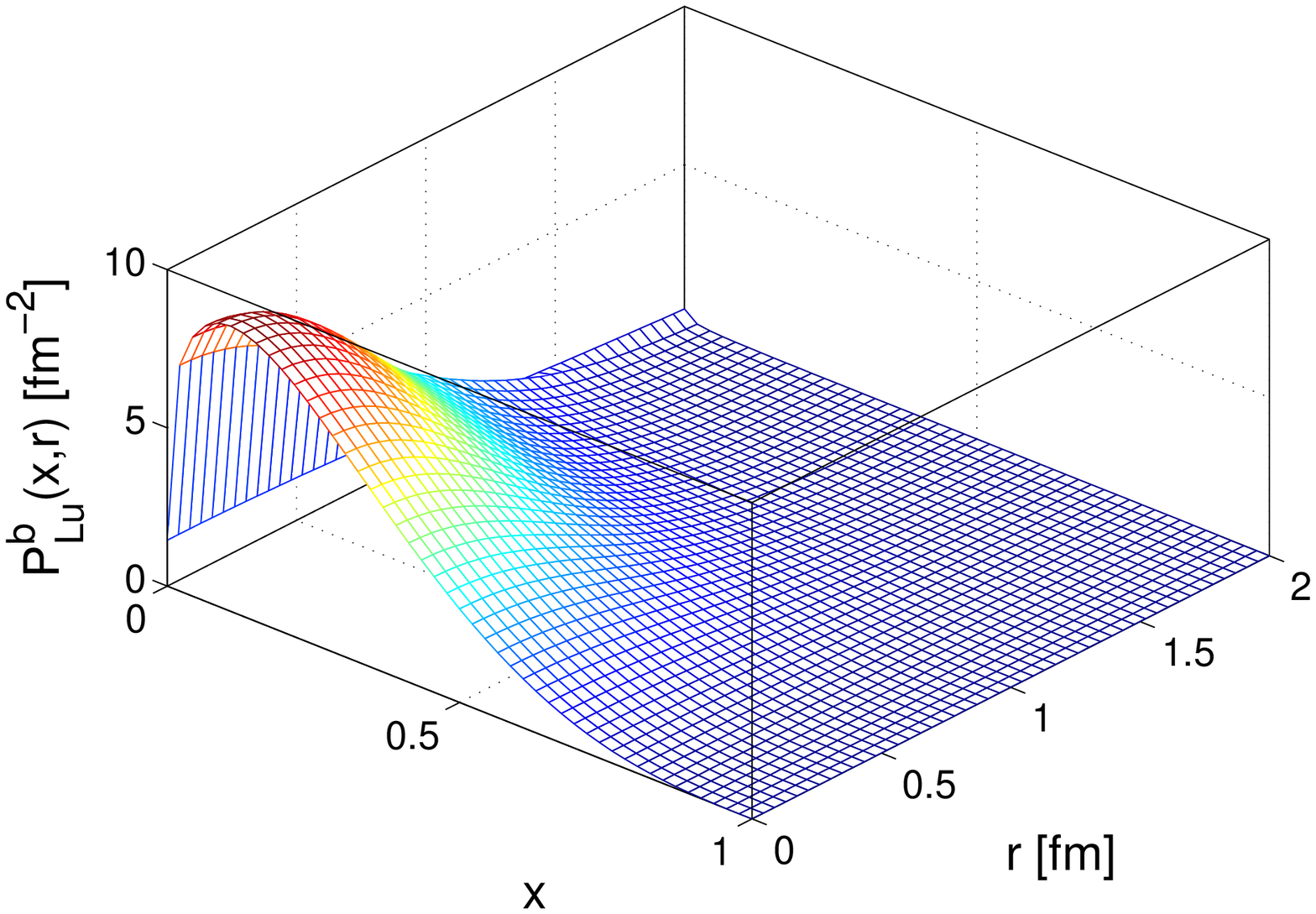}
{(d)}\includegraphics[width=7.5cm,clip]{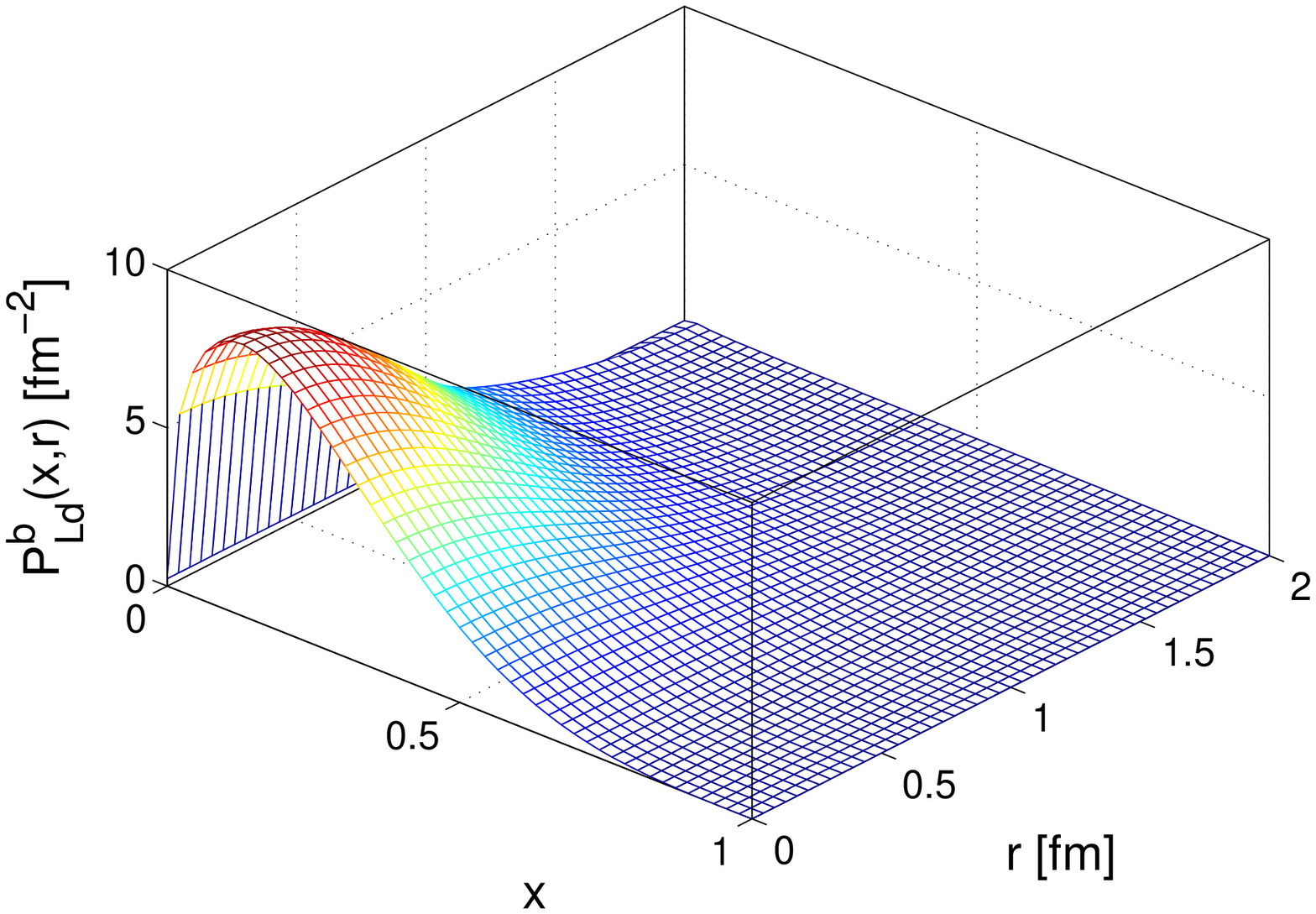}
\end{minipage}
\caption{\label{plot6}(colour online) Longitudinal momentum distributions $P^{f}_{Lq}(x,\bfr)$  and $P^b_{Lq}(x,\bfr)$ as a function of $
\bfr $ and $x$. Left panel: when the struck flavor is $u$ and the diquark is $ud$; right panel:
when the struck flavor is $d$ and the diquark is $uu$.}
\end{figure}
\begin{figure}[htbp]
\begin{minipage}[c]{0.98\textwidth}
{(a)}\includegraphics[width=7.5cm,clip]{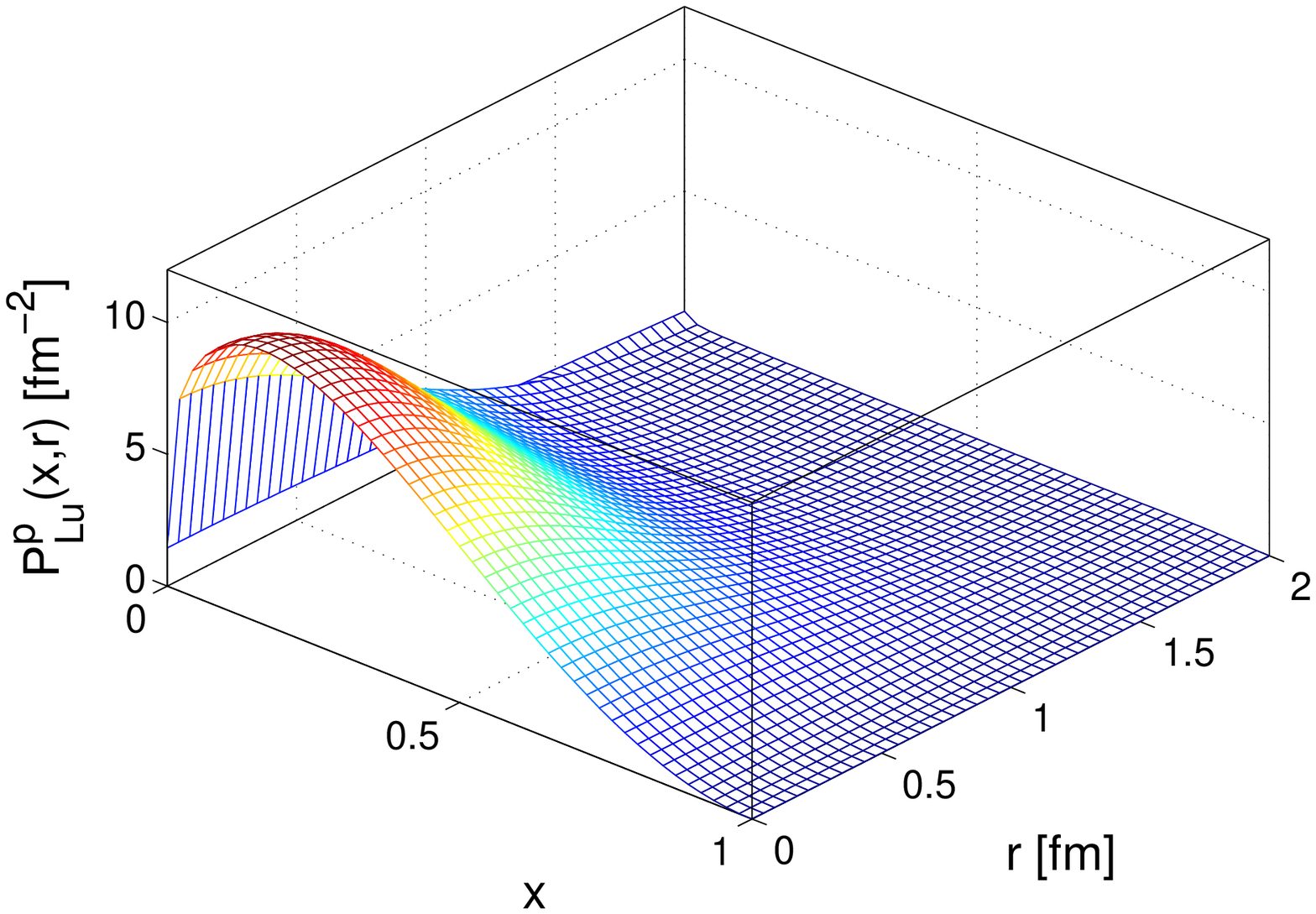}
{(b)}\includegraphics[width=7.5cm,clip]{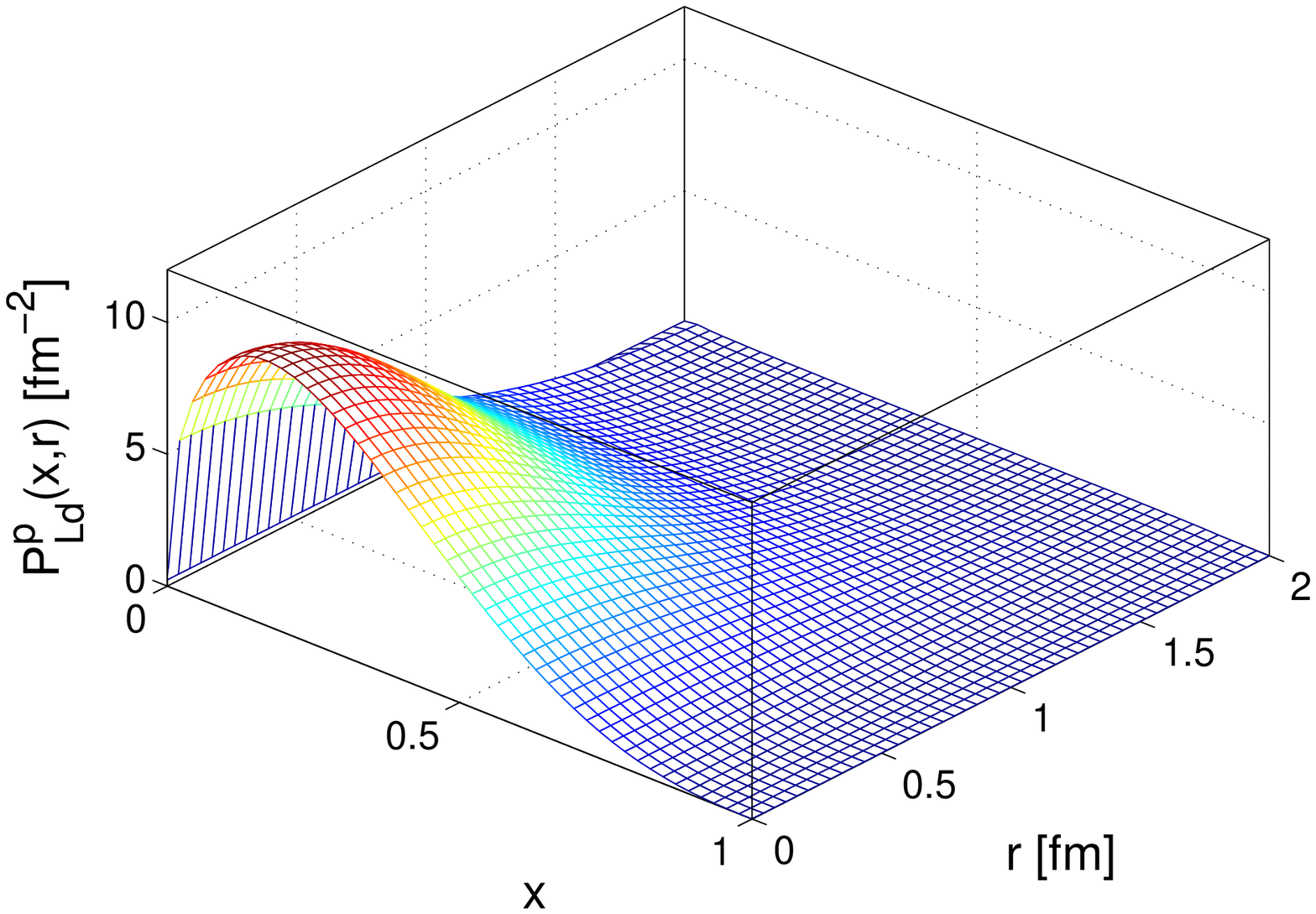}
\end{minipage}
\caption{\label{plot7}(colour online) Longitudinal momentum distributions $P_{Lq}(x,\bfr)$ for proton as a function of $
\bfr $ and $x$, (a) for struck flavor $u$ and (b) for struck flavor $d$. }
\end{figure}

As in the case of the electromagnetic form factors $F_1(q^2)$ and $F_2(q^2)$, the electromagnetic densities were defined from the Fourier transforms, one can in a similar manner interpret the
two-dimensional Fourier transform of gravitational form factor $A(Q^2)$ as the longitudinal
momentum densities in the impact-parameter plane
\cite{abidin08,selyugin,CM5,chakrabarti} as follows
\be
\rho_{L}(b)
&=&\int \frac{d^2\bfq}{(2\pi)^2}A(q^2)e^{i\bfq\cdot\bfb}.
\ee
Using the gravitational form factor in Eq.(\ref{A_FF}) in terms of wave functions and the basic
theorem of convolutions  (see appendix A), we can relate the longitudinal momentum density $\rho_{L}(b)$ in the impact-parameter space with the  density $P_{L}(r)$ in transverse coordinate space. For the quark (fermion), the longitudinal momentum density in the impact-parameter space is given as
\be\label{rel_longi1f}
\rho^f_{Lq}(b)
=\frac{1}{4\pi}\int_0^1dx
\frac{1}{(1-x)^2}P^f_{Lq}\Big(x,\frac{b}{x-1}\Big)=\frac{1}{4\pi}\int_0^1dx
\frac{x}{(1-x)^2}P^q_{f}\Big(x,\frac{b}{x-1}\Big).
\ee
For the diquark (boson), the longitudinal momentum density in the impact-parameter space is given s
\be\label{rel_longi1b}
\rho^b_{Lq}(b)
=\frac{1}{4\pi}\int_0^1dx \frac{1}{x^2}P^b_{Lq}\Big(x,\frac{b}{x}\Big)=\frac{1}{4\pi}\int_0^1dx
\frac{1-x}{x^2}P^q_{f}\Big(x,\frac{b}{x}\Big).
\ee
Therefore, the longitudinal momentum distribution in $\bfb$-space for the nucleons can be expressed as
\be
\rho^N_{Lq}(b)=\rho_{Lq}^f(b)+\rho_{Lq}^b(b).
\ee
The longitudinal distributions in the impact-parameter plane in the light-front
quark-diquark model in AdS/QCD has been investigated in \cite{chakrabarti} but in the present work we extend the calculations to study the longitudinal momentum distribution in transverse coordinate space.
In Fig.\ref{plot5}, we show the longitudinal momentum distributions in transverse coordinate
space in the light-front quark-diquark model using the wave functions modeled by soft-wall
AdS/QCD. The quark (fermionic) and diquark (bosonic) contributions are shown in
Fig.\ref{plot5}(a) and \ref{plot5}(b) for different struck $u$ and $d$ quarks respectively. The
total distribution for proton (quark + diquark) is shown in Fig.\ref{plot5}(c).
The figures show that the diquark contributions are comparatively larger than the quark contributions for both the cases where the struck quarks are $u$ and $d$. Comparing with the phenomenological model, one finds that both the contributions from the quark and diquark to the proton distribution are larger in phenomenological model as compared to that in  the quark-diquark model in AdS/QCD when the struck quark is $u$ quark.  The total distribution i.e. quark$+$diquark distribution should be independent of the struck quark.
For the case of proton, the total distribution  for  $u$ struck quark is slightly larger than that for the  $d$ struck quarkwhereas the difference between these two distributions is quite large in the phenomenological model considered. So, we  can say that the quark-diquark model in AdS/QCD is better compared to the phenomenological model considered here. The distribution $P_{Lq}$ as a function
of $x$ and $r$ for fermionic part as well as the bosonic part considering different struck quark
are plotted in Fig.\ref{plot6}. We find that the fermionic distributions have the peak near the
middle of $x$ whereas for bosonic distributions, the peak shifts to
lower $x$. The fall-off of quarks  distributions  (Fig.\ref{plot6}(a) and Fig.\ref{plot6}(b)) at large $x$ are slower than the diquark
distributions (Fig.\ref{plot6}(c) and Fig.\ref{plot6}(d)). The magnitudes of the diquark distributions are quite large compared to quark
distributions. Thus, the peak of the total (quark + diquark) distributions effectively appears at
lower $x$ but the fall-off the total distributions are little slower than the diquark
distributions as shown in Fig.\ref{plot7}. The magnitudes of the total distributions for
different struck quarks are almost the same.

\section{Summary and Conclusions}\label{summary}
In the present work, we have presented the charge distributions in transverse
coordinate space for $u$ and $d$ quark as well as for the nucleons in a light-front quark-diquark model
where the wave functions are modeled from the solution of two particle wave function in the soft-wall
AdS/QCD correspondence. We have compared the charge distributions with the charge
densities in impact-parameter space. A relation between the distributions in
transverse coordinate
space and the impact-parameter plane has been shown. It has been found that the charge distributions
in coordinate plane are more widely spread compared to the impact-parameter distributions. This
phenomena can be clearly understood from the relation between these two different distributions
which implies that the $\bfr$ is equivalent to $\frac{\bfb}{x-1}$ in $P(\bfr)$. The distributions
$P(x,\bfr)$ for the quarks and the nucleons have the peak at lower $x$ values and gradually decrease in magnitude
with increasing $x$. The distribution for neutron is mostly negative but it has also a
small positive distribution which suggests that the central charge density is negative
covered by a positively charged shell.

We have also studied the longitudinal momentum
distributions for proton and the individual distribution for quark (fermionic) and diquark (bosonic) in the transverse coordinate space from the gravitational form factors. It has been observed that the diquark distributions are larger than the quark distributions but when the quark and diquark contributions are added up, they provide more or less same proton distributions for different struck $u$ and $d$ quarks. $P_{Lq}(x,\bfr)$ for quarks has the peak
near the middle value of $x$ but for diquark the peak is at lower $x$ value. The diquark
distributions fall faster than the quark distributions at large $x$. Since $P_{Lq}(x,\bfr)$ for
diquarks are large compared to that for quarks, the proton distributions have the
peak at small $x$. The distributions however fall slower than the diquark distributions. It is important to mention here that the density in impact parameter space cannot be interpreted as coordinate space density even though both $r$ and $|b_\perp|$ are conjugate to momentum $k$ and momentum transferred $\Delta_\perp$ respectively.
\appendix
\section{Elementary theorems about convolutions and Fourier transforms}\label{appen}
If we have any two functions, $f(\bfk)$ and $g(\bfk)$ such that
\be
f(\bfk)= \int {d^2\bfr} e^ {-i{{\bf k}_\perp}\cdot{{\bf r}_\perp}}
\tilde{f}(\bfr), \quad\quad
g(\bfk)= \int {d^2\bfr} e^{-i{{\bf k}_\perp}\cdot{{\bf r}_\perp}}
\tilde{g}(\bfr),
\ee
where $\tilde{f}(\bfr)$ and $\tilde{g}(\bfr)$ are the Fourier transform of $f(\bfk)$ and
$g(\bfk)$ respectively. Then the form factor which can be written as
\be
G(\bfq)\equiv \int \frac{d^2\bfk}{(2\pi)^2} f^*(\bfk-a\bfq)g(\bfk),
\ee
becomes diagonal in Fourier space and we have
\be
\int \frac{d^2 \bfq}{(2\pi)^2}~
e^{i{{\bf q}_\perp}\cdot{{\bf b}_\perp}} G(\bfq) =
{1\over |a|^2}\ \tilde{f}^*\Big({\bfb\over a}\Big) \tilde{g}\Big({\bfb\over a}\Big).
\ee


\end{document}